\documentclass{article}
\usepackage{url}
\usepackage{amsmath}
\usepackage{amsfonts}
\usepackage{amscd}
\usepackage{graphicx}
\usepackage{color}

\begin{document}

\begin{center}
Scattering with real-time path integrals
\end{center}

\bigskip
\begin{center}
W. N. Polyzou, Ekaterina  Nathanson\\
Proceedings for Few Body 22, Caen, France
\end{center}
\bigskip
\begin{itemize}
\item[] Abstract:Sharp-momentum transition matrix elements for scattering from a
  short-range Gaussian potential are computed using a real-time path
  integral.  The computation is based on a numerical implementation of
  a new interpretation of the path integral as the expectation of a
  potential functional with respect to a complex probability
  distribution on cylinder sets of paths.  The method is closely
  related to a unitary transfer matrix computation.
\end{itemize}

\bigskip
\section{Path integrals and complex probabilities}
Path integrals \cite{Feynman_1}\cite{Feynman_2} provide a means for treating problems in quantum
mechanics that are often difficult to treat by other means.  Normally
path integral calculations are limited to quadratic interactions or
generating perturbation theory. When the time can be made imaginary
they can also be approximated using Monte Carlo \cite{Metropolis} integration.

Most applications are naturally formulated using real time.
Scattering is one such application.  While the computation of
real-time path integrals is more challenging, it is interesting to
explore how far real-time applications can be pushed.  One motivation
is because real-time path integrals represent unitary time evolution,
direct treatments of real-time path integrals have the potential to be
candidates for applications of quantum computing algorithms.

The problem with the interpretation of the real-time path integral as
an integral is that the measure has to be additive on a countable
union of disjoint measurable sets.  When the measure is not positive,
the evaluation of a countable sum of non-positive numbers can be
both infinite or finite, depending on how it is evaluated.

In \cite{Muldowney}\cite{Katya_1}\cite{Katya_2} the Feynman
path integral is reinterpreted
as the expectation of a potential functional
\begin{equation}
  E[F]   \qquad  F[\gamma] := e^{-i \int V(\gamma(t)) dt}
\label{eq.1}  
\end{equation}  
with respect to a complex probability on a space of paths.  It was
shown in \cite{Katya_1}\cite{Katya_2} that this interpretation
results in a global solution of the Schr\"odinger equation.

Classical probabilities on the real line are defined on measurable sets
that are generated by intervals under complements and countable
unions.  These sets are the Lebesgue measurable sets.  A Probability
is a non-negative Lebesgue measurable function that integrates to 1.

The Henstock-Kurzweil integral \cite{Henstock}\cite{bartle} provides
an extension of the Lebesgue integral that is defined similar to a
Riemann integral.  Hesntock developed an alternative probability
theory based on the this integral. Henstock's probability agrees with
classical probability theory when the probability is non-negative,
however he realized that it could be extended to non-positive and
complex probabilities if countable additivity was replaced by finite
additivity.  P. Muldowney \cite{Muldowney} suggested that this
interpretation could also be extended to reinterpret path integrals
not as integrals, but as the expectation of a potential functional
with respect to a complex probability distribution on a space of paths.
J{\o}rgensen and Nathanson verified that this interpretation leads to
approximations that converge to global solutions of the Schr\"odinger
equation.

In one dimension the path-integral representations of the unitary time
evolution operator for a particle of mass $\mu$ in a potential $V$ 
can be expressed as:
\[
\langle x_0 \vert e^{-iHt}  \vert \psi  \rangle =  
\]
\begin{equation}
\lim_{N\to \infty}
({\mu \over 2 \pi i \Delta t})^{N/2}
\int \prod_{i=n}^N dx_n
e^{
i {\mu \over 2\Delta t} (x_{n-1}-x_n)^2
-i V(x_n) \Delta t }
\langle x_N \vert \psi  \rangle.  
\label{eq.2}  
\end{equation}
where $\Delta t := t/N$. This follows from the Trotter product formula
\cite{Simon} and is the standard form of the path integral.
It is expressed as the limit of $N$-dimensional integrals as $N\to \infty$.

The complex probability interpretation arises by expressing the real line
as the union of a finite number of disjoint intervals
\begin{equation}
\mathbb{R} =
\underbrace{(-\infty, x_{1n})}_{I_{0n}}
\underbrace{[x_{1n},x_{2n})}_{I_{1n}} , \cdots , 
\underbrace{[x_{M-1,n},x_{M,n})}_{I_{M-1,n}},
\underbrace{[x_{M,n},\infty)}_{I_M,n} .
\label{eq.3}  
\end{equation}
and replacing the integral over each time slice by a sum of integrals over
each of these intervals
\[
\langle x_0 \vert e^{-iHt}  \vert \psi  \rangle =  
\]
\begin{equation}
\lim_{N\to \infty} 
({\mu \over 2 \pi i \Delta t})^{N/2}
\sum_{m_1 \cdots m_N}
\prod_{n=1}^N  \int_{I_{m_n}} dx_n
e^{
i {\mu \over 2\Delta t} (x_{n-1}-x_n)^2
-i V(x_n) \Delta t }
\langle x_N \vert \psi \rangle.
\label{eq.4}  
\end{equation}

Computationally the intervals should be chosen
so $e^{-i V(x)}$ is approximately constant on each interval.
This also applies to the half-infinite intervals, where
on these intervals, for a scattering
problem with a short range interaction, $e^{-i V(x)} \approx 1$.
When these conditions hold the potential terms can be factored out
of the integral, and evaluated at any point $y_{mn} \in I_{mn}$:
\[
\langle x_0 \vert e^{-iHt}  \vert \psi  \rangle \approx  
\]
\begin{equation}
\lim_{N\to \infty} 
({\mu \over 2 \pi i \Delta t})^{N/2}
\sum_{m_1 \cdots m_N}
\prod_{n=1}^N e^{-i V(y_{mn}) \Delta t } 
\int_{I_{m_n}} dx_n
e^{
i {\mu \over 2\Delta t} (x_{n-1}-x_n)^2}
\langle x_N \vert \psi \rangle.  
\label{eq.5}  
\end{equation}
What remains has the form
\[
\langle x_0 \vert e^{-iHt}  \vert \psi  \rangle \approx
\]
\begin{equation}
\lim_{N\to \infty} 
\sum_{m_1 \cdots m_N}
P(x_0,I_{m1}, \cdots , I_{mN})
e^{-i \sum_n V(y_{mn})\Delta t}
\langle y_{m_N}
\vert \psi \rangle
\label{eq.6}  
\end{equation}
where
\begin{equation}
P(x_0,I_{m1}, \cdots , I_{mN}) =
({\mu \over 2 \pi i \Delta t})^{N/2} 
\prod_{n=1}^N  \int_{I_{m_n}} dx_n
e^{
  i {\mu \over 2\Delta t} (x_{n-1}-x_n)^2 }
\label{eq.6}  
\end{equation}
represents the complex probability that a path passes through $I_{mN}$
at time $t_1$, $I_{mN-1}$ at time $t_2$,  $\dots I_{m1}$ at time $t_N$.
It has all of the properties of a complex probability.  The
approximation converges in the limit that $N$ gets large, the size of
the intervals gets small, and the boundary of the half-infinite
intervals approaches $\pm \infty$.  In principle the intervals and
evaluation points needed for a given accuracy are determined by
the Henstock theory of integration.

This representation has the advantage that, if the intervals are chosen
the same way on each time slice, the potential only needs to be
evaluated at a small number of points where the potential is not zero.

The computational problem is that the number of cylinder sets grows
like $M^N$ in the limit that both $M$ and $N$ become infinite.  
In order to make this computable the complex probability is approximately
factored \cite{polyzou} into sum of products of one-step complex probabilities. They
represent the complex probability of staring at a point in a
given interval and coming out in another interval. This has the property
that sum over all final intervals is 1, since the particle has to
come out in some interval.  The factorization has the form
\begin{equation}
P(x_0,I_{m1},\cdots ,I_{mN}) \approx 
P(x_0,I_{m1}) 
\prod_{n=2}^N P(y_{m({n-1})},I_{nm})
\label{eq.8}  
\end{equation}
where
\begin{equation}
P(y_{m},I_{k}) = 
\sqrt{{1  \over i \pi }}
\int_{\sqrt{{\mu \over 2\Delta t}}(x_k-y_m)}^{\sqrt{{\mu \over 2\Delta t}}(
x_{k+1}-y_m)}
e^{ 
  i \beta^2} d\beta
\label{eq.9}  
\end{equation}
can be evaluated analytically \cite{as}.  With this factorization the
``path integral'' can be approximated by
\[
\langle x_0  \vert e^{-i H t} \vert \psi \rangle \approx
\]
\begin{equation}
\sum_{m1 \cdots mN}
P(x_0,I_{m1}) e^{ -i V(y_{m1}) \Delta t} 
\prod_{n=2}^N P(y_{m({n-1})},I_{nm}) e^{ -i V(y_{mn}) \Delta t}
\langle y_{mN} \vert \psi \rangle . 
\label{eq.10}  
\end{equation} 
This expresses the path integral as the $N$-th power of a complex matrix,
applied to a vector.  Powers of matrices can be efficiently computed.
In addition, because the matrix can be computed analytically,
the matrix elements can be evaluated  on the fly, so it
is not necessary to store large matrices.

The table shows the sum of the one-step probabilities over 5000
intervals for various values of $y$.  The imaginary part
sums to zero with no visible round-off error.

\begin{center}{\bf
Testing complex probabilities (5000 terms)}
\end{center}    
\begin{table}
\begin{center}
\begin{tabular}{l|l|l}
\hline
$y$ & $\sum_n \Re (P_n(x))$ &$\sum_n \Im (P_n(x))$\\
\hline
-25.005001 & p=1.000000e+00 & + i(5.273559e-16) \\
-20.204041 &p=1.000000e+00 &+ i(-1.387779e-16) \\
-15.003001 & p=1.000000e+00 & + i(2.775558e-17) \\
-10.202040 & p=1.000000e+00 &+ i(2.775558e-16) \\
-5.001000  & p=1.000000e+00 &+ i(7.771561e-16) \\
0.200040   &p=1.000000e+00 &+ i(3.608225e-16) \\
5.001000   &p=1.000000e+00 &+ i(1.665335e-16) \\
10.202040  &p=1.000000e+00 &+ i(5.273559e-16) \\
15.003001  &p=1.000000e+00 &+ i(8.049117e-16) \\
20.204041  &p=1.000000e+00 &+ i(1.137979e-15) \\
25.005001  &p=1.000000e+00 &+ i(-2.164935e-15)\\
\hline                                          
\end{tabular}
\caption{\bf sums of complex probabilities}
\end{center}
\end{table} 

The goal of this exercise is to calculate scattering observables.  The
basic observables are functions of on-shell transition matrix elements
\begin{equation}
\langle \mathbf{p}_f \vert T(E_i0^+) \vert \mathbf{p}_i \rangle \approx
\langle \psi_{f0}(0) \vert V e^{-iHt} \vert \psi_{i0}(-t) \rangle
\label{eq.11}  
\end{equation}
where $\langle \mathbf{p} \vert \psi_{f0}(0)\rangle $ and
$\langle \mathbf{p} \vert \psi_{i0}(0)\rangle$  are narrow Gaussian
wave packets centered at $\mathbf{p}_f$ and $\mathbf{p}_i$ with delta function normalizations (i.e. they integrate to 1).  The initial state is evolved
to $-t$ using the free dynamics.
In one dimension the on-shell transition matrix elements are
related to phase shifts and transmission and reflection coefficients by
\begin{equation}
e^{2 i \delta_{\pm} (E)}= 1 - 2 \pi i {m \over \vert p \vert }  
\langle {\pm}p \vert T(E+i0^+) \vert p \rangle
\label{eq.12}  
\end{equation}
\begin{equation}
T=1-{2 \pi i p \over m} \langle p \vert T (E+i0^+) \vert p \rangle
\qquad
R= -{2 \pi i p \over m} \langle -p \vert T (E+i0^+)
\vert p \rangle
\label{eq.13}  
\end{equation}

This method was tested by computing the half-shell scattering matrix
elements for a particle of mass $\mu$ off of a repulsive Gaussian
potential of $-\lambda e^{-(r/r_0)^2}$.  Because this is a
time-dependent computation it is necessary to choose the parameters
carefully.  The time has to be chosen so, after accounting for wave
packet spreading, the wave
packet should be outside of the range of the potential.  The
exhibited calculations are for $\mu=1$, $r_0=1$, $\lambda=5$, $p_i=5$
and $\Delta p=.25$.  The calculations used 120 time steps and 10000
intervals between $\pm 15$.  The results are compared to an exact numerical
calculation.

Figures 1 and 2 show the width of the wave packet and potential.  The
second figure shows that at $t=-3$ the wave packet is out of the range
of the potential, which means that the time limit can be replaced by
an evaluation at $t=3$.  Figures 3 and 4 show the real and imaginary
parts of the half-shell transition matrix elements using the path
integral (short dashes) compared to the exact calculation (long
dashes) of the same quantity.

While this is not the most efficient method to solve this problem, it
demonstrates that real-time path integrals can be used to calculate
scattering observables.  The exhibited calculations used fixed time
steps and interval widths.  There are many possible improvements in
computational efficiency.

The most important observation is that after the one-step
factorization, the method is numerically equivalent to a unitary
transfer-matrix calculation.  The factorization that arises from the
Trotter product formula allows the transfer matrix to be factored into
a product of a potential term and a one step probability matrix.  The
potential term is exactly unitary and the complex one-step probability
matrix is approximately unitary.  Because the calculated quantity is a
matrix element of the product of the potential with a wave operator,
it is only necessary to evolve the system for a finite time in a
finite volume.  These observations are the key to understanding both
the strength and limitations of this method.  A reasonable expectation
is that, with some refinement, this method could be applied to the
same class of problems that can be treated using transfer matrix
methods.

\begin{figure}
\begin{minipage}[t]{.45\linewidth}
\centering
\includegraphics[angle=000,scale=.25]{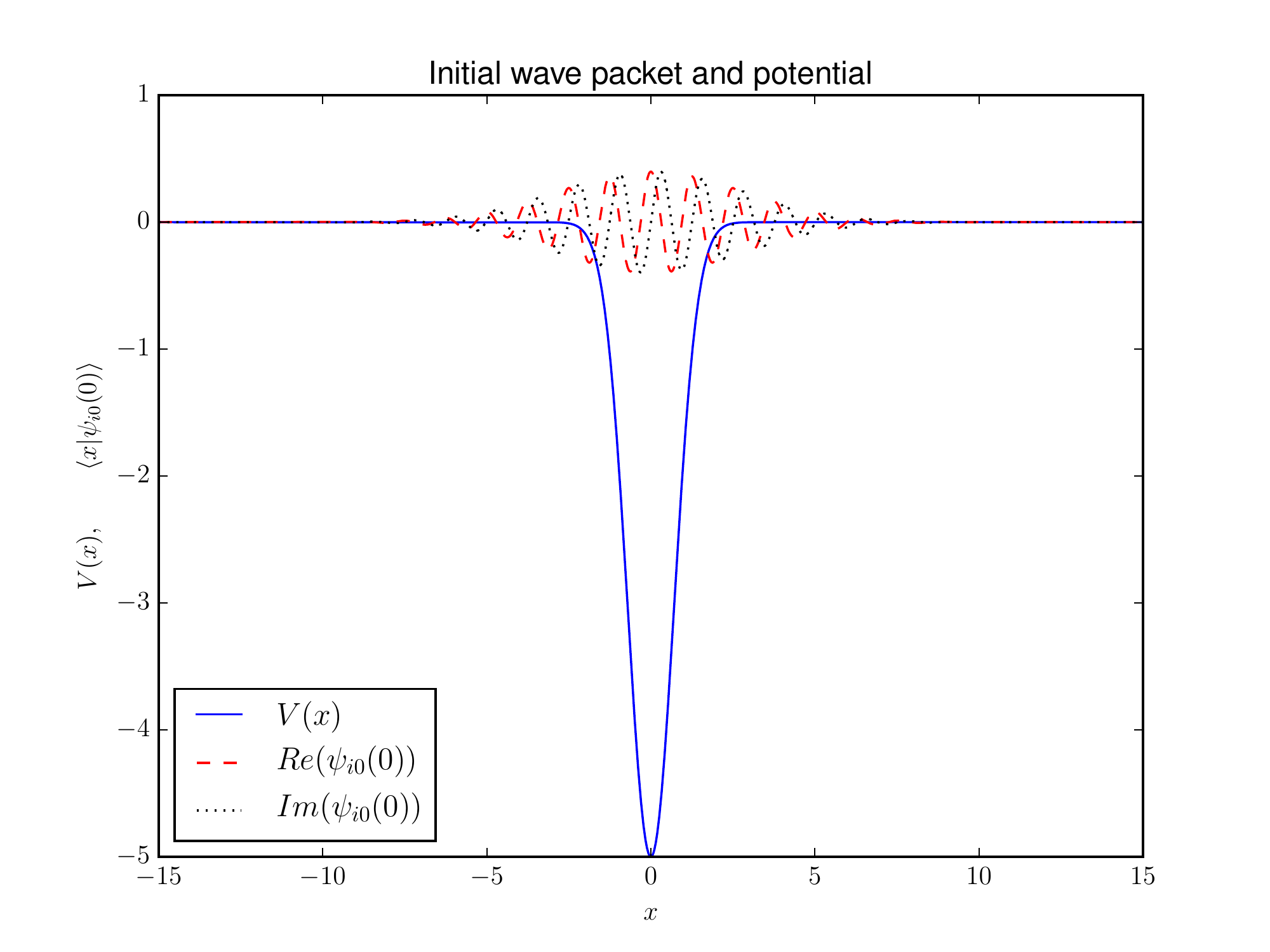}
\caption{\bf V(x),$\langle x \vert \psi_{0i}(0) \rangle$}
\label{fig:1}
\end{minipage}
\begin{minipage}[t]{.45\linewidth}
\centering
\includegraphics[angle=000,scale=.25]{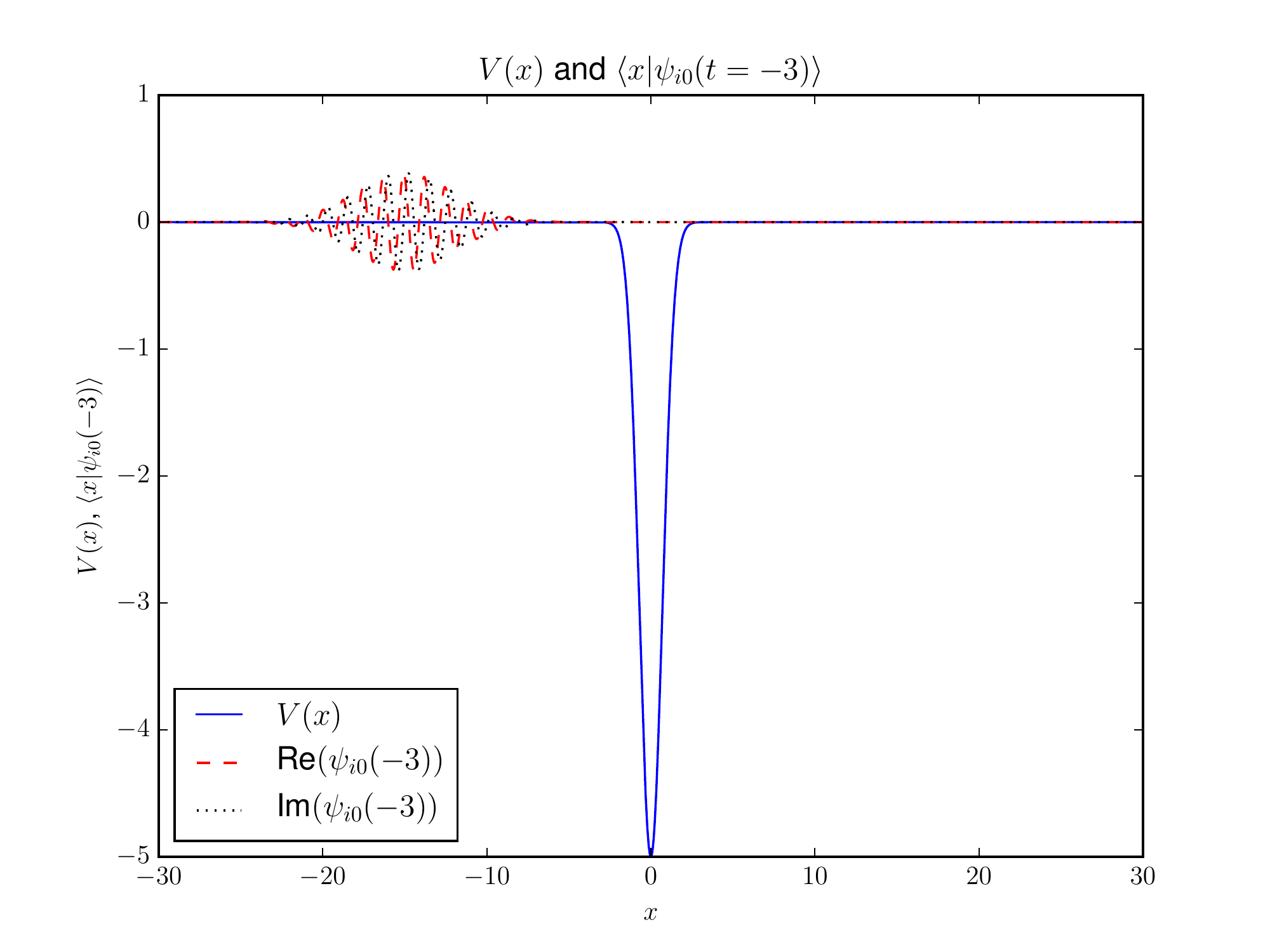}
\caption{\bf V(x),$\langle x \vert \psi_{0i}(-3) \rangle$}
\label{fig:2}
\end{minipage}
\end{figure}

\begin{figure}
\begin{minipage}[t]{.45\linewidth}
\centering
\includegraphics[angle=000,scale=.25]{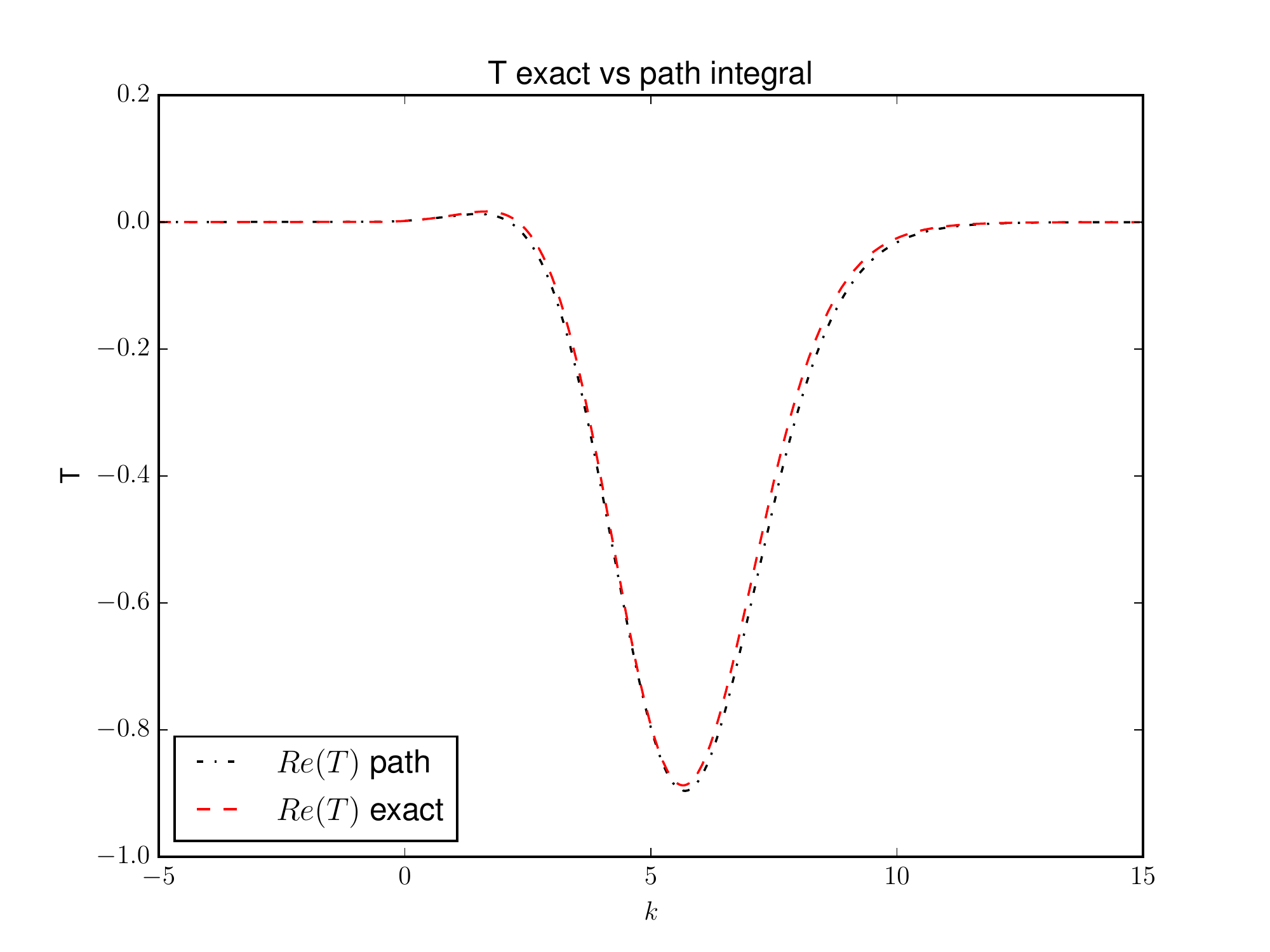}
\caption{\bf Re(T) exact vs path integral}
\label{fig:17}
\end{minipage}
\begin{minipage}[t]{.45\linewidth}
\centering
\includegraphics[angle=000,scale=.25]{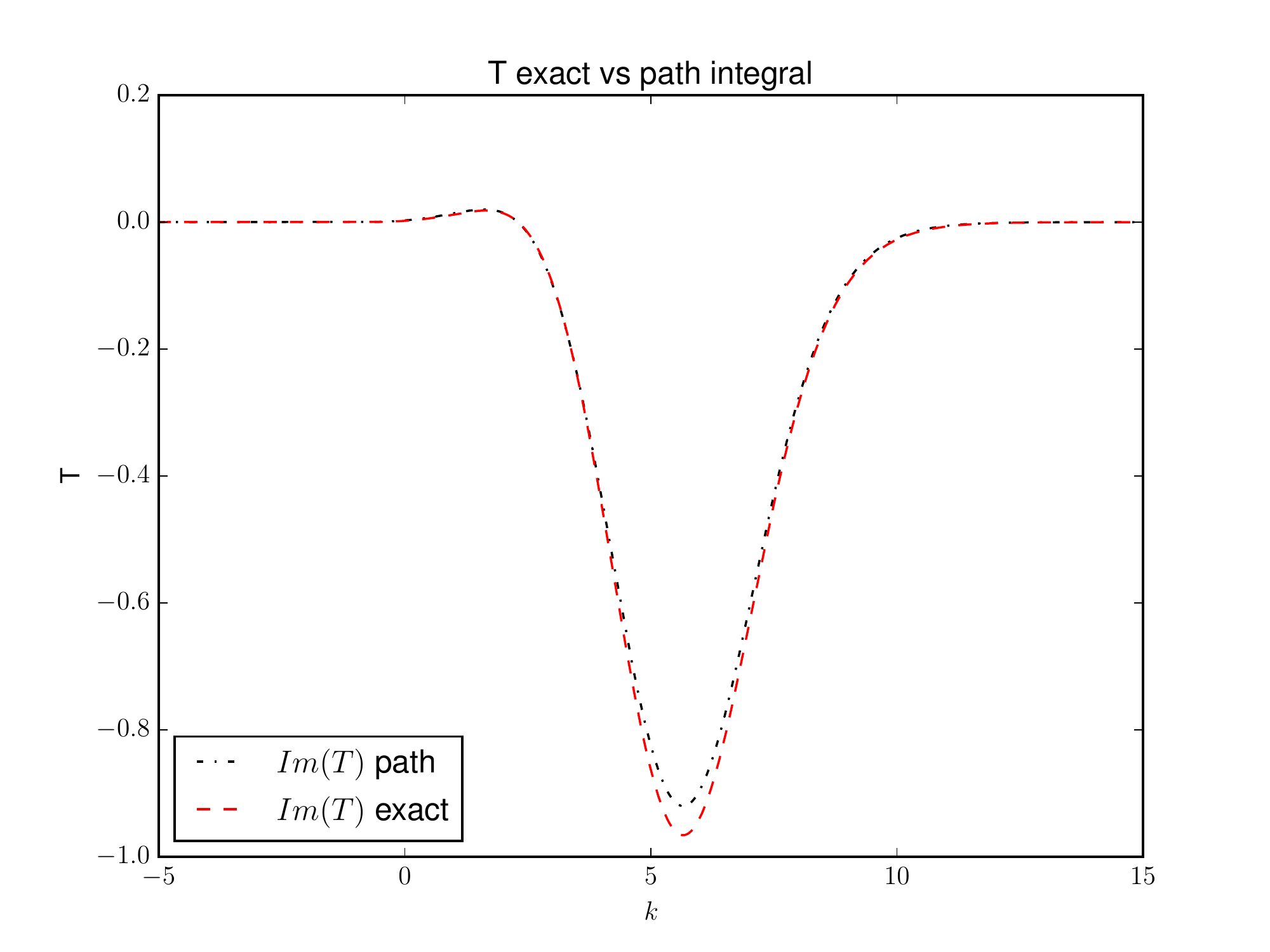}
\caption{\bf Im(T) exact vs. path integral}
\label{fig:18}
\end{minipage}
\end{figure}

This work supported by the U.S. Department of Energy, Office of Science,
Grant number: DE-SC0016457.


\begin{thebibliography}{14}

\bibitem{Feynman_1}
Feynman, R.:
Space-time approach to non-relativistic quantum mechanics,
Reviews of Modern Physics,2,367(1948)

\bibitem{Feynman_2}
Feynman, R. and Hibbs, A. R.:
{\it Quantum Mechanics and Path Integrals},
McGraw-Hill, USA (1965)
  
\bibitem{Metropolis}
Metropolis, N. and Ulam S.:
The Monte Carlo method,
Journal of the American Statistical Association,
44,335(1949)

\bibitem{Muldowney}
Muldowney, P.:
{\it A Modern Theory of Random Variation},
Wiley, NJ, (2012)
  
\bibitem{Katya_1}
Nathanson, Ekaterina S. and  J{\o}rgensen, Palle, E.T.:
A global solution to the Schr\o:dinger equation: From Henstock to Feynman,
J. Math. Phys. 56, 092102(2015)

\bibitem{Katya_2}
Nathanson, Ekaterina S.:
Path integration with non-positive distributions and
applications to the Schr\"odinger equation,
University of Iowa Thesis,
(2015)
  
\bibitem{Henstock}
Henstock, R.:
{\it Theory of Integration},
Butterworths, London, (1963)

\bibitem{bartle}
Bartle, R. G.:
{\it A modern theory of integration},
American Mathematical Society, Providence, RI (2001)

\bibitem{Simon}
Reed, M. and Simon, B.:
{\it Methods of Modern Mathematical Physics},
Academic Press, San Diego, p 295(1980)

\bibitem{polyzou}  
Polyzou, W. N. and Nathanson, E. S.:
Scattering using real-time path integrals,
arxiv:1712.00046,(2018)

\bibitem{as}
Abramowitz, M. and Stegun, I.:
{\it Handbook of Mathematical Functions},
National Bureau of Standards, Washington, DC, (1964)

\end{thebibliography}
%
\end{document}